\documentclass[fleqn,twoside]{article}
\usepackage{espcrc2}

\usepackage{graphicx}
\usepackage[figuresright]{rotating}

\newcommand{\AmS}{{\protect\the\textfont2
  A\kern-.1667em\lower.5ex\hbox{M}\kern-.125emS}}
\def\nin{\noindent}
\def\beq{\begin{equation}}
\def\eeq{\end{equation}}
\def\bea{\begin{eqnarray}}
\def\eea{\end{eqnarray}}

\newcommand{\strich}[1]{#1  \! \! \! \slash}
\newcommand{\cO}{{\cal O}}

\title{$\tau^-\to\pi^-\gamma\nu_\tau$ as a source of background on the Lepton Flavour Violating decay $\tau^-\to\mu^-\gamma$}

 \author{
 Zhi-Hui Guo\address{College of Physics and Information Engineering, Hebei Normal University, Shijiazhuang 050016, People's Republic of China. } ,
 Pablo Roig\address{Laboratoire de Physique Th\'eorique.
Universit\'e de Paris-Sud XI, B\^atiment 210, 91405. Orsay cedex, France.}
\thanks{I wish to congratulate the organizing Committee for the brilliant organization of the TAU2010 Conference. I have benefited from useful comments by Michel Davier and
George Lafferty and a clarifying discussion with Alberto Lusiani, all of them concerning the possibility of detecting the photon in $\tau\to\pi\gamma\nu_\tau$
at the B-factories and how to fix a related realistic energy cut on $E_\gamma$. I also thank Gabriel L\'opez Castro for pointing out a complementary analysis
that can help the detection of this mode and checks of our description. I am indebted to Zbigniew Was and Tomasz Przedzinski for providing me with data 
representing the photon spectrum due to background from $\tau\to\pi\nu_\tau$ where radiation is simulated with PHOTOS. This work has been supported in part by the
EU MRTN-CT-2006-035482 (FLAVIAnet), by MEC (Spain) under grant FPA2007-60323, by the Spanish Consolider-Ingenio 2010 Programme CPAN (CSD2007-00042).}}

\begin{document}

\begin{abstract}
\noindent We calculate the decay of $\tau^-\to\pi^-\gamma\nu_\tau$ in the framework of Resonance Chiral Theory ($R\chi T$). 
By demanding the high energy constraints from QCD on the related form factors, we could  predict the various physical 
observables of $\tau^-\to\pi^-\gamma\nu_\tau$ without any free parameter. Our results show that for a realistic cut on the photon energy (around $100$ MeV) this mode gives a branching ratio of roughly $0.1\%$ that should have already
been detected at the heavy-flavour factories. Another interesting subject we have studied based on our calculation of the decay $\tau^-\to\pi^-\gamma\nu_\tau$ is the experimental background estimation of the lepton flavour violation process $\tau^- \to \mu^- \gamma$. 
We point out that although the description of radiation that PHOTOS provides -which has been used by BaBar and Belle collaborations to estimate this source 
of background- is in excellent agreement with the theoretical expectations in the low energy region in $\tau^-\to\pi^-\gamma\nu_\tau$
decay, this is not the case in the high energy region, precisely where it is easier that this decay mimics the process $\tau^-\to\mu^-\gamma$.

\end{abstract}

\maketitle






\section{Hadronic decays of the $\tau$ lepton} \label{Haddecs}
\begin{figure}[h!]
\centering
\includegraphics[scale=0.3]{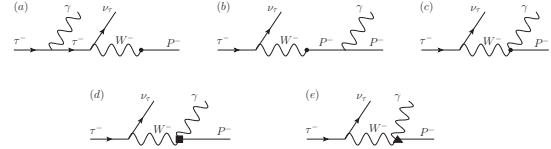}
\caption{\small{Feynman diagrams for the different kinds of contributions to the radiative decays of the tau including a pion, as explained in the main text. The
 dot indicates the hadronization of the $QCD$ currents. The solid square represents the $SD$ contribution mediated by the vector current and the solid
 triangle the $SD$ contribution via the axial-vector current.}}
\label{diagrams general decomposition amplitude in radiative decays tau with one meson}
\end{figure}

\nin The decays of the $\tau$ lepton into hadrons are a very interesting subject in Particle Physics that has been a very active area of research during the
last 25 years \cite{Tau}. The $\tau$ is privileged because its mass allows it to decay hadronically while being a lepton, which means a clean place to scrutinize
hadronization effects at low-energies, where QCD becomes nonperturbative. One can exploit this advantage to a great extent through its inclusive decays \cite{InclusiveTau}.
However we will focus in what follows in one of its exclusive decays.\\
The decay amplitude for the one meson radiative decays of the $\tau$ includes an internal bremsstrahlung (IB) component, that is given by QED, and thus can
be calculated unambiguously to any desired order in perturbation theory. In addition, one has the structure dependent (SD) part, dominated by the effects
of the strong interaction and the interference (INT) between both. Lorentz symmetry determines that there are two independent structures, the so-called vector
(V) and axial-vector (A) form factors that encode our lack of knowledge of the precise mechanism responsible for hadronization.\\
We consider the $\tau^-(p_\tau) \to \nu_\tau(q) \pi^-(p) \gamma(k)\,$ process \footnote{The decay $\tau^-(p_\tau) \to \nu_\tau(q) K^-(p) \gamma(k)\,$, which is
related to it via chiral symmetry, is discussed in Ref.~\cite{Guo:2010dv}, in which this proceeding is based.}. The kinematics of this decay is equivalent
to that of the radiative pion decay \cite{Brown:1964zza}. We will use $t := (p_\tau - q)^2 = (k + p)^2$. In complete analogy to that case \cite{Bae67}, the
matrix element for the decay of $\tau^- \to \pi^- \gamma \nu_\tau$ can be written as the sum of four contributions:
\begin{equation} \label{mtot}
   \mathcal{M} \left[\tau^- \to \nu_\tau \pi^- \gamma\right]
   = \mathcal{M}_{IB_\tau} + \mathcal{M}_{IB_\pi} + \mathcal{M}_{V} + \mathcal{M}_{A}\,,\,
\end{equation}
with \footnote{Definitions and conventions differ with respect to Ref.~\cite{Decker:1993ut} ($DF$), in which we have detected~\cite{Guo:2010dv} several typoes.}
{\footnotesize
\begin{eqnarray} \label{Gral_IB}
& &  i \mathcal{M}_{IB_{\tau}} =  G_F V_{ud} e F_\pi p_\mu \epsilon_\nu(k) L^{\mu \nu}\,, \nonumber \\
& &  i \mathcal{M}_{IB_{\pi}} = G_F V_{ud} e F_\pi \epsilon^\nu(k)\left( \frac{2p_\nu (k + p)_\mu}
      {m_\pi^2 - t} + g_{\mu\nu} \right) L^\mu, \nonumber \\
& &  i \mathcal{M}_V = i G_F V_{ud} e F_V(t) \varepsilon_{\mu \nu \rho \sigma}
     \epsilon^\nu(k)  k^\rho p^\sigma  L^\mu,  \nonumber \\
& &  i \mathcal{M}_{A} = G_F V_{ud} e F_A(t) \epsilon^\nu(k) \left[(t-m_\pi^2) g_{\mu\nu}-2p_\nu k_\mu \right] L^\mu, \nonumber
\end{eqnarray}
}
where $e$ is the electric charge of the positron and $\epsilon_\nu$ is the polarization vector of the photon. $F_V(t)$ and $F_A(t)$ are the so called 
$SD$ form factors. Finally $L^\mu$ and $L^{\mu\nu}$ are lepton currents defined by
{\footnotesize
\begin{eqnarray}
   L^\mu & = & \bar{u}_{\nu_\tau}(q) \gamma^\mu (1-\gamma_5) u_\tau(p_\tau)\,, \nonumber \\
   L^{\mu \nu} & = & \bar{u}_{\nu_\tau}(q) \gamma^\mu (1-\gamma_5)
\frac{\strich{k} - \strich{p}_\tau
      - M_\tau}{(k - p_\tau)^2 - M_\tau^2} \gamma^\nu u_\tau(p_\tau)\,.\nonumber
\end{eqnarray}
}
The notation introduced for the amplitudes describes the four kinds of contributions: $\mathcal{M}_{IB_{\tau}}$ is the bremsstrahlung off
the tau, (Figure \ref{diagrams general decomposition amplitude in radiative decays tau with one meson}(a)); $\mathcal{M}_{IB_{\pi}}$ is the sum of
the bremsstrahlung off the $\pi$ (Figure \ref{diagrams general decomposition amplitude in radiative decays tau with one meson}(b)), and the
seagull diagram (Figure \ref{diagrams general decomposition amplitude in radiative decays tau with one meson}(c)); $\mathcal{M}_{V}$ is the
$SD$ vector contribution (Figure \ref{diagrams general decomposition amplitude in radiative decays tau with one meson}(d)) and $\mathcal{M}_{A}$
the $SD$ axial-vector contribution (Figure \ref{diagrams general decomposition amplitude in radiative decays tau with one meson}(e)).
Our ignorance of the exact mechanism of hadronization is parametrized in terms of the two form factors $F_A(t)$ and $F_V$(t). In fact, these form
factors are the same functions of the momentum transfer $t$ as those in the radiative pion decay, the only difference being that $t$ now varies
from $0$ up to $M_\tau^2$ rather than just up to $m_\pi^2$.
The dimensionless variables
\begin{equation}
 x := \frac{2 p_\tau \cdot k}{M_\tau^2}\,,\quad \quad y := \frac{2 p_\tau \cdot p}{M_\tau^2}\,,
\end{equation}
allow to measure the photon and pion energies in units of $\frac{M_\tau}{2}$ in the tau rest frame. Their kinematical boundaries are given by
\begin{eqnarray}
   0 \leq x \leq & 1 - r_\pi^2\,,\quad 1 - x + \frac{r_\pi^2}{1 -x} \leq y \leq 1 + r_\pi^2\,,\,
\end{eqnarray}
where
\begin{equation} \label{r_pi2}
 r_\pi^2:= \left( \frac{m_\pi}{M_\tau} \right)^2 \sim 0\mathrm{.}006 \ll 1\,,
\end{equation}
so that the photon spectrum will extend up to $\sim0.994\frac{M_\tau}{2}$. In the decay $\tau\to\mu\gamma$ the photon energy would be fixed to
$\frac{M_\tau}{2}\left(1-\frac{m_\mu^2}{M_\tau^2}\right)\sim0.996\frac{M_\tau}{2}$ so one could expect some contamination of the decay $\tau^- \to \pi^- \gamma \nu_\tau$
 to $\tau\to\mu\gamma$, with the $\pi$ misidentified as a $\mu$. The invariant mass of the $\pi$-$\gamma$ system will vary in the range
$r_\pi^2 M_\tau^2\leq t \leq M_\tau^2$ when the other independent kinematical variable is taken as $x=x(z)$, with $z= t/M_{\tau}^2$ .
\section{Theoretical setting} \label{Theory}
\begin{figure}[ht]
\begin{center}
\includegraphics[scale=0.3]{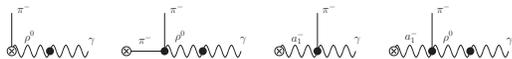}
\caption{\small{Axial-vector current contributions to $\tau^-\rightarrow  \pi^- \gamma \nu_\tau$.} \label{fig.pi.a}}
\end{center}
\end{figure}

\begin{figure}[ht]
\begin{center}
\includegraphics[scale=0.3]{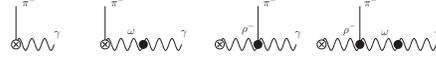}
\caption{\small{Vector current contributions to $\tau^-\rightarrow  \pi^- \gamma \nu_\tau$.} \label{fig.pi.v}}
\end{center}
\end{figure}

Tau lepton decays probe QCD in its non-perturbative regime where standard expansions in powers of the coupling constant are no longer applicable. However, the chiral symmetry of the massless theory allows to build an effective field theory dual
to QCD in the light-meson sector, Chiral Perturbation Theory ($\chi PT$) \cite{ChPT}. This will not be enough, since $t$ can reach $M_\tau^2$, which is far beyond the region of applicability
of $\chi PT$. Still, the low-energy region of semileptonic tau decays is well described by the $\chi PT$ results~\cite{Colangelo:1996hs}. Then,
one should envisage a way of enlarging the range of applicability of $\chi PT$ while respecting its low-energy behaviour. The limit of a large number of
colors ($N_C$) in QCD \cite{Nc} is a useful tool in the development of Resonance Chiral Theory ($R \chi T$) \cite{RChT}, a Lagrangian formulation including the resonances as
dynamical degrees of freedom that preserves assymptotic QCD properties \cite{BrodskyLepage}, \cite{Floratos:1978jb}. Remarkably, $R \chi T$ was capable of predicting
the low-energy constants (LECs) of $\chi PT$ in terms of resonance masses and couplings. The phenomenological application of the theory (in the $N_C\to\infty$ limit)
to study two-\cite{hh} and three-meson decays of the $\tau$ \cite{hhh} has been successful and several Green-functions \cite{VVP}, \cite{GFs} and associated form factors
have been studied within it over the years.\\
The leading action of the $R \chi T$ Lagrangian includes the $\cO(p^2)$ $\chi PT$ in the even-intrinsic parity sector and the leading $\cO(p^4)$ $\chi PT$
(given by the Wess-Zumino-Witten term \cite{WZW}) in the odd-intrinsic parity sector. Higher-order pieces in the chiral expansion are assumed to be generated
by the integration of the resonances (this was checked to occur at $\cO(p^4)$ for the $L_i$ couplings of $R \chi T$ \cite{RChT}).\\
Next, one adds all pieces including resonances (R) and chiral tensors with low enough chiral order ($\chi^{(n)}$) to not violate high-energy conditions or to force fine-tuned cancellations
among the relevant couplings to fulfil the short distance constraints. In the odd-intrinsic-parity sector, that contributes to the vector form factor, this amounts to include all terms of $R \chi^{(4)}$
and $RR \chi^{(2)}$. For the even-intrinsic-parity operators that are contributing to the axial-vector form factors, these are the terms of $R\chi^{(2)}$. Since previous
analysis showed the relevance of the $RR\chi^{(2)}$ with negative intrinsic parity we will consider them here, as well. All mentioned pieces of the Lagrangian
can be found in Ref.~\cite{Guo:2010dv}, as well as the expressions for the $SD$ form factors in $\tau\to\pi\gamma\nu_\tau$ decays. In Figs. \ref{fig.pi.a}
and \ref{fig.pi.v} the Feynman diagrams contributing to these form factors are displayed. The thick dots represent strong vertices.
\\
We provide the resonances with an adequate energy-dependent width obtained consistently within $R\chi T$ \cite{widths} and consider only one resonance multiplet
per set of quantum numbers (the contribution of spin-zero resonances is suppressed by conservation laws and the fact that there is only one meson in the final
state). Since this process has not been measured yet, one should try the simplest possible description (in order to be predictive) that can eventually be
completed if the data require so.\\
Next we require a Brodsky-Lepage \cite{BrodskyLepage} behaviour to the vector form factor and demand that the axial-vector form factor satisfies a once
subtracted dispersion relation. This results in constraints among the Lagrangian couplings. Noteworthily, the relations found in the $\tau\to \pi/K \gamma \nu_\tau$
decays are all consistent with those obtained in the $\tau$ decays into two and three mesons. There is only one relation that differs with respect to the study of
the Vector-Vector-Pseudoscalar Green's function \cite{VVP} for a coupling whose impact is very mild and the discrepancy is less than $10\%$.
\section{Phenomenology in $\tau^-\to\pi^-\gamma\nu_\tau$ decays} \label{Pheno}
\begin{figure}[h!]
\begin{center}
\includegraphics[scale=0.3,angle=-90]{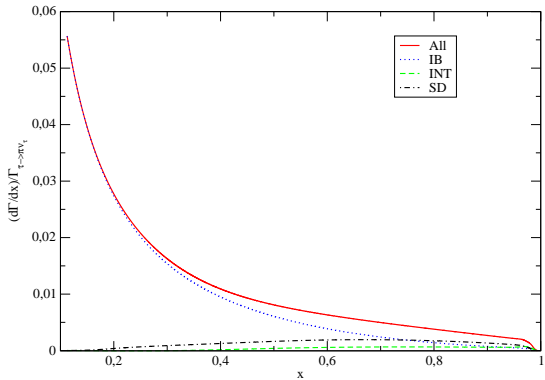}
\caption{\small{Differential decay width of the process $\tau^-\rightarrow  \pi^- \gamma \nu_\tau$ including all contributions as a function of $x$.} \label{All}}
\end{center}
\end{figure}

\begin{figure}[h!]
\begin{center}
\includegraphics[scale=0.3,angle=-90]{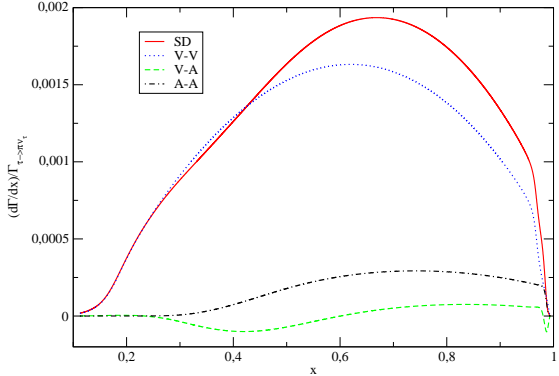}
\caption{\small{Differential decay width of the process $\tau^-\rightarrow  \pi^- \gamma \nu_\tau$ including only the structure dependent contributions as a function of $x$.} \label{SD}}
\end{center}
\end{figure}

First we have assessed the importance of the model independent contributions, meaning the $IB$ described by QED and the Wess-Zumino-Witten contribution to the vector
form factor, that is determined from QCD. We have thus switched off the remaining contributions to the vector form factor and the whole axial-vector form factor
in this first step. We find that for a cut on the photon energy of $100$ MeV, it amounts to $0.9\%$ of the non-radiative decay, namely a branching fraction of
$0.1\%$. Next, we include also the model dependent contributions, obtained as described in Sect.~\ref{Theory}.\\
In Fig.~\ref{All} we show the differential decay width of the process $\tau^-\rightarrow  \pi^- \gamma \nu_\tau$ including all contributions as a function of $x$, i.e. the photon energy in the tau rest frame and in Fig.~\ref{SD} we display the $SD$ contributions, that are enhanced near the endpoint region, where the process
$\tau^-\to\pi^-\gamma\nu_\tau$ can contaminate the decay $\tau^-\to\mu^-\gamma$.
\section{Comparison with PHOTOS} \label{Comparison}
\begin{figure}[h!]
\begin{center}
\includegraphics[scale=0.265,angle=-90]{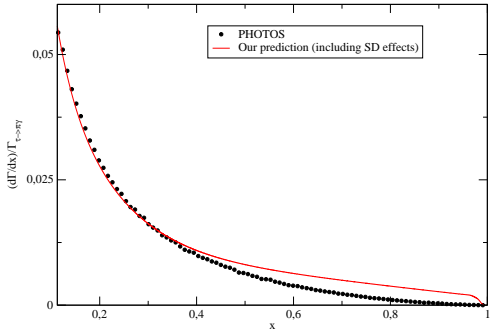}
\caption{\small{Comparison of our prediction with PHOTOS for the differential decay width of the process $\tau^-\rightarrow  \pi^- \gamma \nu_\tau$ including
all contributions (in our case) as a function of $x$. In both cases $E_\pi$ is integrated over all its range.} \label{Comparison1}}
\end{center}
\end{figure}

\begin{figure}[h!]
\begin{center}
\includegraphics[scale=0.265,angle=-90]{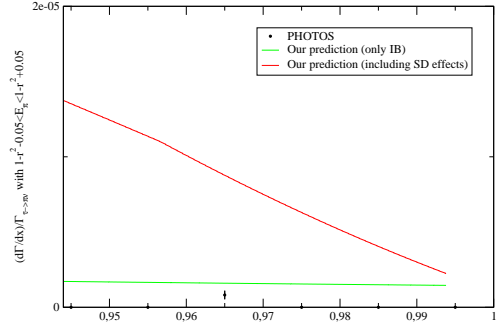}
\caption{\small{Comparison of our prediction with PHOTOS for the differential decay width of the process $\tau^-\rightarrow  \pi^- \gamma \nu_\tau$ including
all contributions (in our case) as a function of $x$. In both cases $E_\pi$ is integrated over the range $1-r_\pi^2-0.05<y<1-r_\pi^2+0.05$.} \label{Comparison2}}
\end{center}
\end{figure}

\nin In the previous section, we have shown that for a realistic cut on the photon energy (around $100$ MeV) this mode gives a branching ratio of roughly 
$0.1\%$ that should have already been detected at the heavy-flavour factories.  Notwithstanding, this decay mode has not been measured yet. This does not 
mean that the detection of soft photons at B-factories is not as good as estimated, but that the splitting of the radiative and non-radiative pion decay of 
the tau was not considered a priority.\\
According to the last report by the Belle~\cite{Belle} Collaboration \footnote{Although the BaBar report \cite{BaBar} is not that detailed, we assume similar 
figures to hold in their case.}, the main source of background in $\tau^-\to\mu^-\gamma$ searches is the process $\tau^-\to\pi^-\gamma\nu_\tau$, where the 
pion is miss-identified as a muon. This decay represents up to $80\%$ of all background and the pion fake probability to be detected as a muon is $0.8\%$. 
However, the most important contribution to this background would come from radiation off the $e^+e^-$ pair and not off the $\pi$ or $\tau$. The background 
related to the decay $\tau^-\to\pi^-\gamma\nu_\tau$ is estimated using PHOTOS \cite{PHOTOS} that only incorporates the model independent contributions from 
QED to $\tau^-\to\pi^-\gamma\nu_\tau$, i.e. the IB parts in Eq.(\ref{mtot}). For low photon energies, this decay is dominated by the IB parts and is detected 
as $\tau^-\to\pi^-\nu_\tau$, since the photon cannot be resolved. However, we have seen in Fig. \ref{All}, that for large photon energies the dominant 
contribution is given by the $SD$ terms. Indeed, it is precisely near the endpoint of the photon decay spectrum where it is easier that the miss-ID happens, 
since in the two body-decay $\tau^-\to\mu^-\gamma$, $E_\gamma\sim E_\mu\sim M_\tau/2$, and the correction is $1\mp0.004$, while in $\tau^-\to\pi^-\gamma\nu_\tau$ 
the photon energy spectrum extends essentially to the same value and the correction here is given by $1-r_\pi^2$, see Eq. (\ref{r_pi2}).\\
We compare our prediction for the photon spectrum in $\tau^-\to\pi^-\gamma\nu_\tau$ with what is obtained with PHOTOS \cite{PHOTOS} and MC-TESTER \cite{MC-TESTER} 
runned with fixed first order only (similar results are obtained with exponentiation on), as it is displayed in Fig. \ref{Comparison1}.\\
In order to assess the impact of the $SD$ effects in that background estimation, one needs to compare PHOTOS and our prediction for the case where the $\pi$ 
and $\gamma$ energies reconstruct the $\tau$ mass up to 9 MeV of difference. We do this in Fig. \ref{Comparison2}, where this missing mass is distributed evenly 
between the $\pi$ and the $\gamma$ and only the relevant region near the endpoint is displayed in order to better appreciate the differences. The PHOTOS simulation 
was obtained with 100Mevents generated for  $\tau^-\to\pi^-(\gamma)\nu_\tau$ where $~2.9\%$ corresponds to the radiative decay. We see that in the last six 
bins there is only one event. We can estimate -rather conservatively- the difference between our prediction and PHOTOS by taking the integral over these last six 
bins, where we will estimate the PHOTOS contribution by our prediction including only $IB$. This gives a ratio of $\sim5$, as the underestimation of background 
due to the $SD$ effects we have studied.\\
Nevertheless, we stress that the total decay width is mildly affected by the $SD$ parts so that the global description provided by PHOTOS is good. 
For example, for the total decay width one has
\begin{eqnarray}
\Gamma_{\tau \to \pi \gamma \nu_\tau}^{ALL} /\Gamma_{\tau \to \pi \gamma \nu_\tau}^{IB} (E_\gamma=100{\rm MeV}) &=& 1.1 \,.\nonumber
\end{eqnarray}
Our estimations (factor of 5 more background due to the considered decay that obtained with PHOTOS) would imply that this source of background could affect 
the present upper bounds in Refs.~\cite{Belle},~\cite{BaBar}:$4.5\times10^{-8}$ at $90\%$ $CL$, making them even stronger. The new version \cite{Olga} of 
hadronic currents in TAUOLA \cite{TAUOLA} and the inclusion on PHOTOS of our current in the line of Ref. \cite{Nanava:2009vg} will be the ideal tools to 
estimate reliably this background.\\
\section{Conclusions and Outlook} \label{Conclusions}
We have analyzed the decay $\tau^-\to\pi^-\gamma\nu_\tau$. 
It turns out
essential to measure the photon energy spectrum near the endpoint in order to assess the real background affecting the lepton flavour violating decay $\tau^-\to\mu^-\gamma$, 
and the upper bound for its branching ratio. Moreover, it would be an interesting tool for lepton universality tests through the ratio 
$\Gamma(\tau^-\to\pi^-\gamma\nu_\tau)$/$\Gamma(\pi^-\to\mu^-\gamma\bar{\nu}_\mu)$. To our view this decay channel should no longer be regarded as attached 
to the $\tau^-\to\pi^-\nu_\tau$ measurement for energetic enough photons which would allow experimental resolution. Therefore, it would become a golden 
mode to be searched for by the (super)B and tau-charm factories in the near future. 


 

\begin{thebibliography}{00}

\bibitem{Tau}
  A.~Pich,
  ``QCD Tests From Tau Decay Data,''
 M.~Davier, A.~Hocker and Z.~Zhang,
 Rev.\ Mod.\ Phys.\  {\bf 78}, 1043 (2006).
 J.~Portol\'es,
 Nucl.\ Phys.\ Proc.\ Suppl.\  {\bf 169} (2007) 3.
 A.~Pich,
 Nucl.\ Phys.\ Proc.\ Suppl.\  {\bf 181-182} (2008) 300.
  S.~Actis {\it et al.},
  Eur.\ Phys.\ J.\  C {\bf 66}, 585 (2010).

\bibitem{InclusiveTau}
A. Pich, these proceedings; M. Jamin, these proceedings.

\bibitem{Guo:2010dv}
 Z.~H.~Guo and P.~Roig,
  arXiv:1009.2542 [hep-ph].

\bibitem{Brown:1964zza}
 S.~G.~Brown and S.~A.~Bludman,
 Phys.\ Rev.\  {\bf 136} (1964) B1160.

\bibitem{Bae67}
 P. de Baenst and J. Pestieau,
 Nuovo Cimento 53A, 407 (1968).

\bibitem{Decker:1993ut}
 R.~Decker and M.~Finkemeier,
 Phys.\ Rev.\  D {\bf 48} (1993) 4203
 [Addendum-ibid.\  D {\bf 50} (1994) 7079].

\bibitem{ChPT}
S.~Weinberg,
Physica A {\bf 96} (1979) 327.
J.~Gasser and H.~Leutwyler,
Annals Phys.\  {\bf 158} (1984) 142;
Nucl.\ Phys.\  B {\bf 250} (1985) 465.

\bibitem{Colangelo:1996hs}
G.~Colangelo, M.~Finkemeier and R.~Urech,
Phys.\ Rev.\  D {\bf 54} (1996) 4403.

\bibitem{Nc}
G.~'t Hooft,
Nucl.\ Phys.\  B {\bf 72} (1974) 461;
{\bf 75} (1974) 461.
E.~Witten,
Nucl.\ Phys.\  B {\bf 160} (1979) 57.

\bibitem{RChT}
G.~Ecker, J.~Gasser, A.~Pich and E.~de Rafael,
Nucl.\ Phys.\  B {\bf 321} (1989) 311.
G.~Ecker, J.~Gasser, H.~Leutwyler, A.~Pich and E.~de Rafael,
Phys.\ Lett.\  B {\bf 223} (1989) 425.

\bibitem{BrodskyLepage}
S.~J.~Brodsky and G.~R.~Farrar,
Phys.\ Rev.\ Lett.\  {\bf 31} (1973) 1153.
G.~P.~Lepage and S.~J.~Brodsky,
Phys.\ Rev.\  D {\bf 22} (1980) 2157.

\bibitem{Floratos:1978jb}
 E.~G.~Floratos, S.~Narison and E.~de Rafael,
 Nucl.\ Phys.\  B {\bf 155} (1979) 115.

\bibitem{hh}
F.~Guerrero and A.~Pich,
Phys.\ Lett.\  B {\bf 412} (1997) 382.
M.~Jamin, A.~Pich and J.~Portol\'es,
Phys.\ Lett.\  B {\bf 640} (2006) 176;
Phys.\ Lett.\  B {\bf 664} (2008) 78.
  Z.~H.~Guo,
  Phys.\ Rev.\  D {\bf 78} (2008) 033004.

\bibitem{hhh}
 D.~G.~Dumm, A.~Pich and J.~Portol\'es,
 Phys.\ Rev.\  D {\bf 69} (2004) 073002.
 D.~G.~Dumm, P.~Roig, A.~Pich and J.~Portol\'es,
 Phys.\ Lett.\  B {\bf 685} (2010) 158.
 D.~G.~Dumm, P.~Roig, A.~Pich and J.~Portol\'es,
 Phys.\ Rev.\  D {\bf 81} (2010) 034031.
 D.~G.~Dumm, P.~Roig and A.~Pich, to appear.

\bibitem{VVP}
P.~D.~Ruiz-Femen\'{\i}a, A.~Pich and J.~Portol\'es,
JHEP {\bf 0307} (2003) 003.

\bibitem{GFs}
M.~Knecht and A.~Nyffeler,
Eur.\ Phys.\ J.\  C {\bf 21} (2001) 659.
G.~Amor\'os, S.~Noguera and J.~Portol\'es,
Eur.\ Phys.\ J.\  C {\bf 27} (2003) 243.
V.~Cirigliano, G.~Ecker, M.~Eidem\"uller, A.~Pich and J.~Portol\'es,
Phys.\ Lett.\  B {\bf 596} (2004) 96.
V.~Cirigliano, G.~Ecker, M.~Eidem\"uller, R.~Kaiser, A.~Pich and J.~Portol\'es,
JHEP {\bf 0504} (2005) 006;
 Nucl.\ Phys.\  B {\bf 753} (2006) 139.

\bibitem{WZW}
 J.~Wess and B.~Zumino,
 Phys.\ Lett.\  B {\bf 37} (1971) 95.
 E.~Witten,
 Nucl.\ Phys.\  B {\bf 223} (1983) 422.

\bibitem{widths}
D.~G.~Dumm, A.~Pich and J.~Portol\'es,
Phys.\ Rev.\  D {\bf 62} (2000) 054014.

\bibitem{Belle}
K.~Hayaska {\it et al.}, Phys.\ Lett. \ B {\bf 666} (2008) 16; K.~Inami, these Proceeedings.

\bibitem{BaBar}
B.~Aubert {\it et al.}, Phys.\ Rev. \ Lett. {\bf104} (2010) 021802; A.~Cervelli, these Proceeedings.

\bibitem{PHOTOS}
  P.~Golonka and Z.~Was,
  Eur.\ Phys.\ J.\  C {\bf 45} (2006) 97;
  Eur.\ Phys.\ J.\  C {\bf 50} (2007) 53;
  G.~Nanava and Z.~Was,
  Eur.\ Phys.\ J.\  C {\bf 51} (2007) 569.

\bibitem{MC-TESTER}
 P.~Golonka, T.~Pierzchala and Z.~Was,
  Comput.\ Phys.\ Commun.\  {\bf 157} (2004) 39;
 N.~Davidson, P.~Golonka, T.~Przedzinski and Z.~Was,
  arXiv:0812.3215 [hep-ph].

\bibitem{Olga}
P.~Roig, O.~Shekhovtsova and Z.~Was. Work in progress.

\bibitem{TAUOLA}
 S.~Jadach, Z.~Was, R.~Decker and J.~H.~K\"uhn,
 Comput.\ Phys.\ Commun.\  {\bf 76} (1993) 361.
 Z.~Was, N.~Davidson, G.~Nanava, T.~Przedzinski, E.~Richter-Was, P.~Roig, O.~Shekhovtsova, Q.~Xu, these Proceedings.

\bibitem{Nanava:2009vg}
  G.~Nanava, Q.~Xu and Z.~Was,
  arXiv:0906.4052 [hep-ph].
 \end{thebibliography}
\end{document}